\documentclass[conference]{IEEEtran}
\IEEEoverridecommandlockouts
\usepackage{balance}
\usepackage{cite}
\usepackage{amsmath,amssymb,amsfonts}
\usepackage{algorithm}
\usepackage[noend]{algpseudocode}
\usepackage{booktabs}
\usepackage[mathscr]{euscript}
\usepackage{graphicx}
\usepackage{hyperref}
\usepackage{inputenc}[utf8]
\usepackage{listings}
\usepackage[frozencache=true,cachedir=.]{minted}
\usepackage{theorem}    
\usepackage{textcomp}
\usepackage{tikz}
\usetikzlibrary{matrix}
\usetikzlibrary{arrows}
\usepackage{xcolor}
\def\BibTeX{{\rm B\kern-.05em{\sc i\kern-.025em b}\kern-.08em
    T\kern-.1667em\lower.7ex\hbox{E}\kern-.125emX}}

\definecolor{rpcolor}{rgb}{0.05,0.4,0.1}
\definecolor{vvcolor}{rgb}{0.1,0.1,0.7}
\definecolor{kscolor}{rgb}{0.9,0.1,0.1}

\newcommand{\code}{\texttt}
\newcommand{\tech}{bonsai fuzzing} \newcommand{\Tech}{Bonsai fuzzing} \newcommand{\TECH}{Bonsai Fuzzing}

\begin{document}

\title{Growing A Test Corpus with \TECH{}
}
\author{\IEEEauthorblockN{Vasudev Vikram}
\IEEEauthorblockA{\textit{University of California, Berkeley}\\
Berkeley, CA, USA \\
vasumv@berkeley.edu}
\and
\IEEEauthorblockN{Rohan Padhye}
\IEEEauthorblockA{\textit{Carnegie Mellon University}\\
Pittsburgh, PA, USA \\
rohanpadhye@cmu.edu}
\and
\IEEEauthorblockN{Koushik Sen}
\IEEEauthorblockA{\textit{University of California, Berkeley}\\
Berkeley, CA, USA \\
ksen@cs.berkeley.edu}
}

\maketitle

\begin{abstract}
This paper presents a coverage-guided grammar-based fuzzing technique for automatically synthesizing a corpus of concise test inputs. We walk-through a case study of a compiler designed for education and the corresponding problem of generating meaningful test cases to provide to students. The prior state-of-the-art solution is a combination of fuzzing and test-case reduction techniques such as variants of delta-debugging. Our key insight is that instead of attempting to minimize convoluted fuzzer-generated test inputs, we can instead grow concise test inputs {by construction} using a form of iterative deepening. We call this approach \emph{\tech{}}. Experimental results show that \tech{} can generate test corpora having inputs that are $16$--$45\%$ smaller in size on average as compared to a fuzz-then-reduce approach, while achieving approximately the same code coverage and fault-detection capability.
\end{abstract} 

\section{Introduction}
\label{sec:intro}

This paper describes a new technique for automatically generating a concise corpus of test inputs having a well-defined syntax and non-trivial semantics (e.g. for a compiler).

This project originated when the authors were faced with the task of generating a test corpus for use in an undergraduate compilers course. The course project targets the ChocoPy programming language~\cite{Padhye19-chocopy}. ChocoPy is a statically typed subset of Python, designed specifically for education. In a ChocoPy-based course, students are expected to build a compiler in Java that statically checks and then translates ChocoPy programs to RISC-V assembly. Student projects can be autograded by comparing their compilers' output at various stages---parser, type checker, and code generator---with the corresponding output produced by a reference implementation. 
When starting their project, students are provided with a suite of ChocoPy test programs and the autograder, which together serve as a partial executable specification. This workflow simulates \emph{test-driven development}, while also enabling students to continuously get feedback about their progress. For instructors, writing test cases to validate every language feature is a tedious task; we wanted to \emph{automatically} synthesize such a test corpus. This paper describes the technique we developed for this purpose. In particular, we focus on the problem of automatically generating test cases that exercise the typechecker, since generating well-typed programs is known to be a difficult problem~\cite{Yang11, Palka11, Dewey15, Fetscher15}.

This task presents two conflicting challenges: (1) the generated test suite must be \emph{comprehensive} in covering various semantics of the language, including corner cases; (2) the test suite must be \emph{concise} and readable; in particular, each test case should be small in size so that test failures can guide students towards identifying which feature was incorrectly implemented. The conflict is apparent from previous work~\cite{Arcuri12} which indicates that automated test generation for covering difficult program branches works better with larger test cases.

Much work has been done on automatically generating concise and comprehensive unit test suites~\cite{Pacheco07, Fraser10, Fraser12}. However, this work mainly focuses on generating test {code} as sequences of method calls while minimizing the {number of test cases} or {size of the entire test suite}. Our goal is to generate non-trivial test \emph{inputs} (e.g. strings) while minimizing the  \emph{individual size of each test case}, on average. This is because our conciseness goals are related to readability and debuggability~\cite{Lei05, Leitner07} rather than reducing the cost of test execution~\cite{Harman10}.

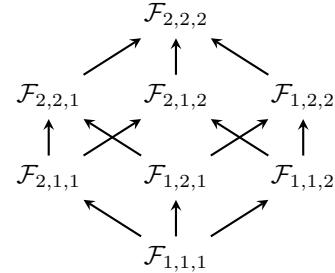
\begin{figure}
    \centering
        \begin{tikzpicture}
        \matrix (a) [matrix of math nodes, column sep=0.6cm, row sep=0.5cm]{
        & \mathcal{F}_{2,2,2} &\\
        \mathcal{F}_{2,2,1} & \mathcal{F}_{2,1,2} & \mathcal{F}_{1,2,2}\\
        \mathcal{F}_{2,1,1} & \mathcal{F}_{1,2,1} & \mathcal{F}_{1,1,2}\\
        & \mathcal{F}_{1,1,1} &\\};
        
        \foreach \i/\j in {2-1/1-2, 2-2/1-2, 2-3/1-2, 3-1/2-1, 3-1/2-2, 3-2/2-1, 3-2/2-3,3-3/2-2, 3-3/2-3, 4-2/3-1, 4-2/3-2, 4-2/3-3}
            \path[>=stealth, thick, ->] (a-\i) edge (a-\j);
        \end{tikzpicture}
    \caption{An example \tech{} architecture---A lattice of coverage-guided size-bounded grammar-based fuzzers $\mathcal{F}_{m,n,d}$, ordered by three size bounds on the syntax of the test cases they produce: number of unique identifiers $m$, maximum sequence length $n$, and maximum nesting depth $d$. Test cases flow along directed edges: the inputs generated by each fuzzer are used as the seed inputs to its successors. The result of \tech{} is the corpus produced by the top-most element.}
    \label{fig:lattice}
\vspace{-0.5cm}
\end{figure}

The state-of-the-art in concise automatic test case generation for structured input domains such as compilers is as follows: first, perform some form of random fuzzing~\cite{Miller90, Godefroid20} to automatically discover unexpected or coverage-increasing inputs. Then, perform test-case reduction~\cite{Zeller02, Misherghi06} on every fuzzer-saved input in order to find a corresponding (locally) minimal input that causes the test program to exhibit the same behavior. For example, CSmith~\cite{Yang11} and C-Reduce~\cite{Regehr12} complement each other by respectively generating and minimizing C programs for automated testing of C compilers.

By their very nature, fuzzer-generated inputs exercise program features chaotically. This can make isolating the most significant features of a fuzzer-saved test input challenging both for humans and for minimization algorithms. Further, to make the test-case minimization problem tractable, algorithms such as delta-debugging~\cite{Zeller02} and its variants perform only local optimization. 

In this paper, we present \emph{\tech{}}, a technique for automatically generating a concise and comprehensive test corpus of structurally complex inputs. Our key insight is that instead of reducing large convoluted inputs that exercise many program features at once, we can grow a concise test corpus \emph{bottom-up}. \Tech{} generates small inputs \emph{by construction} in an iterative evolutionary algorithm: the first round generates tiny trivial inputs and then each successive round generates inputs of slightly larger size by mutating inputs saved in a previous round. In particular, we first define a procedure to sample syntactically valid inputs from a grammar specification using bounds on the number of identifiers, linear repetitions, and nested expansions in the resulting derivation trees. We then define a partial order over coverage-guided bounded grammar fuzzers (CBGFs). For any given desired size bound, this partial order results in a lattice of CBGFs. A corpus produced by each CBGF in this lattice is used as a set of seed inputs for all successive CBGFs. The bottom element of the lattice has minimum size bound---a fuzzer with no good seed inputs---and the top element has the maximum desirable size bound---a fuzzer that produces the final test corpus. Fig.~\ref{fig:lattice} visualizes \tech{} for a given size bound.

Experimental results on the ChocoPy typechecker indicate that \tech{} produces inputs that are 42.5\% smaller than those produced by the fuzz-then-reduce approach, while retaining 98.5\% of coverage and 98.8\% of the mutation score.

Although we developed \tech{} to solve a specific problem related to the use of ChocoPy, the technique is more generally applicable. We report results of applying \tech{} to the Google Closure Compiler, which optimizes JavaScript programs: \Tech{} results in test corpora that are 16.5\% smaller on average than those produced by the conventional fuzz-then-reduce approach, while achieving approximately the same code coverage. We have made a replication package publicly available at \textcolor{blue}{ \href{https://github.com/vasumv/bonsai-fuzzing}{https://github.com/vasumv/bonsai-fuzzing}}. \section{Background and Motivation}
\label{sec:background-motivation}

\subsection{ChocoPy}
\label{sec:background-chocopy}

ChocoPy~\cite{Padhye19-chocopy} is a statically typed subset of Python 3.6. It uses Python's type annotation syntax, but enforces static type checking. Figures~\ref{fig:chocopy-ex1} and \ref{fig:chocopy-ex2} show examples of well-typed ChocoPy programs demonstrating a variety of language features borrowed from Python.

\begin{figure}
    \begin{minted}
    [
    framesep=2mm,
    xleftmargin=5mm,
    baselinestretch=1.2,
    fontsize=\footnotesize,
    linenos
    ]   
    {python}
def is_zero(items:[int], idx:int) -> bool:
    val:int = 0
    val = items[idx]
    return val == 0
idx:int = 1
print(is_zero([1, 0, 1, 0, 1], idx))
    \end{minted}
    \caption{ChocoPy Program illustrating functions, variables,
and static typing. Prints \code{True} when executed.}
    \label{fig:chocopy-ex1}
\end{figure}

\begin{figure}
    \begin{minted}
    [
    fontsize=\footnotesize,
    xleftmargin=5mm,
    linenos,
    escapeinside=||
    ]   
    {python}
class A(object):
    x:int = 1
    def setx(self: "A", y:int):
        self.x = y
    def equals(self: "A", y:int) -> bool:
        return self.x == y
a:A = None
a = A()
if True:
    if a.equals(0):
        a.setx(3)
print(a.x)
    \end{minted}
    \caption{ChocoPy Program illustrating classes, methods, objects, and conditional statements.}
    \label{fig:chocopy-ex2}
    \vspace{-0.2cm}
\end{figure}

ChocoPy is primarily used in undergraduate compilers corpus-sizes. For the programming assignments, students implement a Java-based ChocoPy compiler, whose output is compared against that produced by a publicly available reference compiler. \emph{Autograding} is supported by the ChocoPy infrastructure out-of-the-box.
In this paper, we are interested in specifying and autograding the type-checking component of student-developed ChocoPy compilers: on semantically valid programs, their output is expected to match the type-annotated ASTs (in JSON format) with those produced by the reference compiler; on invalid programs, error messages and corresponding line numbers are compared. A comprehensive test suite therefore consists of both valid and invalid ChocoPy programs that exercise various aspects of the ChocoPy typing rules.

\tikzstyle{block}=[draw, minimum size=2em]
\tikzstyle{arrow} = [pin edge={to-,thin,black}]
\tikzset{every picture/.style={/utils/exec={\rmfamily}}}

\subsection{Problem Definition}
\label{sec:motivation-problem}

Our high-level goal in this paper is to \emph{automatically} synthesize test cases for the ChocoPy typechecker that are not only \emph{comprehensive} but also \emph{concise}. We expand on these primary goals as follows:

\begin{enumerate}
    \item \emph{Automatic}: Manual test creation is cumbersome and error-prone. Further, we want to have the option of quickly adding and removing language features in ChocoPy to evolve its scope. We therefore want a mechanism to automatically generate a test corpus, given only a syntax definition (i.e., a grammar) and a reference compiler implementation.
    \item \emph{Comprehensiveness}: We want the automatically generated test corpus to have high \emph{code coverage} and \emph{fault-detection ability}. We focus on optimizing for branch coverage in the reference compiler and also measure mutation scores where applicable.
    \item \emph{Conciseness}: We want to generate minimal test cases that exercise various features in the reference compiler. We focus on optimizing for individual test-case size, though we also measure the size of the test corpus in number of test cases.
    \item \emph{Semantic Validity}: We want a high fraction of semantically valid programs. Although invalid programs are necessary to cover specific aspects (e.g. error messages) of the typechecker, we prefer generally prefer test cases to be semantically valid as they are more representative examples of language features. 
\end{enumerate}

Finally, it only makes sense to invest in automation if our efforts can be applied to more than one testing target. We therefore also add a secondary goal:

\begin{enumerate}
    \setcounter{enumi}{4}
    \item \emph{Generalizable}: We would like the technique to generalize to at least one other testing target.
\end{enumerate}

On the surface, this seems like a standard automated testing problem. Why do we need a new technique? We next briefly discuss prior work in the context of our application goals and why we felt the need to develop a novel solution.
 
\section{Challenges with Prior Solutions}
\label{sec:related}

\subsection{Systematic Testing}

Since our goal is to generate \emph{concise} test cases, a  natural approach to consider is simply enumerating a bounded space of inputs or program behaviors.

\subsubsection{Bounded Exhaustive Testing}

Tools such as Korat~\cite{Boyapati02}, TestEra~\cite{Khurshid04}, ASTGen~\cite{Daniel07} and UDITA~\cite{Gligoric10} perform \emph{bounded exhaustive testing}: inputs of a bounded size are generated systematically, while employing various optimizations. These tools have been effective at generating test suites for data structure libraries, for powering automatic refactoring tools, etc. Unfortunately, the input space of a ChocoPy compiler is too large to be enumerated exhaustively. The number of unique syntactically valid programs, with at most one user-defined identifier, up to two statements per block, and a maximum block/expression nesting depth of two, is more than the estimated number of atoms in the universe: about $10^{85}$.

\subsubsection{Input Structure}

Since we know the ChocoPy syntax, we can consider systematically enumerating \emph{k-paths}~\cite{Havrikov19} within the ChocoPy grammar. This approach yields minimal programs corresponding to each unique $k$-length path (from root to leaf) in valid syntax trees. This works really well for generating parser tests; however, it is not ideal for exercising the type checking and semantic analysis logic of a compiler. For example, a minimal well-typed ChocoPy program that contains a valid method-call invocation requires several syntax sub-trees to ensure valid class definition, valid method definition inside the class, valid instantiation of an object of that class, and a valid method call on this object. This semantic feature cannot therefore be defined as a linear $k$-path for any $k$. 

\subsubsection{Symbolic Execution}

Instead of enumerating the input space, tools such as JPF-SE~\cite{Anand07} systematically explore the space of \emph{program paths} using symbolic execution~\cite{King76}. With the use of constraint solvers, one could potentially generate a comprehensive test suite that covers a diverse set of program paths of bounded size (assuming that execution path length correlates with input size).
However, the number of program paths to explore grows \emph{exponentially} with the number of branches encountered during execution~\cite{Cadar13}. Even on the small ChocoPy program in Fig.~\ref{fig:chocopy-ex1}, the reference compiler executes {12,274 conditional jumps} and {5,132 virtual method calls}. Exhaustive symbolic execution is therefore not a practical solution even for bounded input sizes.

\subsection{Fuzz Testing}

Random test generation is an established technique for sampling complex input spaces with the hope of discovering unexpected corner cases. The term \emph{fuzz testing} (or simply \emph{fuzzing}) is generally used for techniques that randomly generate \emph{test inputs}~\cite{Miller90, Godefroid20} as opposed to test code~\cite{Pacheco07, Fraser10, Fraser12}. Fuzzing is mainly used for discovering security vulnerabilities.

There are two main challenges in using fuzz testing tools for test corpus generation. First, generating a \emph{comprehensive} test corpus for a compiler requires generating a diverse set of inputs satisfying complex constraints (e.g. programs should be well-typed). We therefore consider several variants of fuzzing that address effective state-space exploration. Second, fuzzer-generated inputs are often notoriously large and unreadable. We thus consider some advances in making test inputs \emph{concise} and \emph{semantically valid}.

\subsubsection{Coverage-Guided Fuzzing}

One one extreme end, coverage-guided fuzzing (CGF) uses no knowledge of the input domain; instead, it instruments programs under test to analyze their run-time behavior. CGF evolves a corpus of test inputs with the goal of maximizing code coverage. 
The process starts with developer-provided or randomly generated \emph{seed inputs}. New inputs are created by performing \emph{random mutations} on seed inputs (e.g. randomly inserting, modifying, or deleting bytes at randomly chosen locations). Inputs that cause the test program to cover previously uncovered code are added to the set of seeds. The process repeats until a time budget expires.
AFL~\cite{AFL} and LibFuzzer~\cite{libFuzzer} are popular CGF tools for finding bugs in programs that parse binary data (e.g. media players and network protocol implementations). When applied to the ChocoPy compiler, these tools are useful for generating tests for the frontend; indeed, AFL helped discover some dormant bugs in the reference parser. However, these tools are ineffective at generating comprehensive tests for the type checker. In a preliminary experiment, we found that less than 0.01\% of AFL-generated inputs were valid ChocoPy programs. This is unsurprising because random byte-level mutations rarely lead to the generation of inputs that can satisfy syntactic and semantic constraints.

\subsubsection{Specialized Compiler Fuzzing}

On the other extreme end, a highly precise compiler fuzzer can be developed by incorporating the syntax \emph{and} semantics of the language in the input generation process itself. For example, CSmith~\cite{Yang11} generates C programs while avoiding undefined behavior, Pa\l{}ka et al.~\cite{Palka11} generate well-typed lambda terms for testing the Glasgow Haskell Compiler, and Dewey et al.~\cite{Dewey15} use constraint logic programming to test the Rust type-checker. Such specialized compiler fuzzers require quite a bit effort to develop, and do not meet our secondary criteria of being generally applicable to multiple testing targets.

\subsubsection{Grammar-based Fuzzing}

Between these extremes, grammar-based fuzzers offer an acceptable middle ground. Using only a declarative specification of a compiler's input grammar---which is often readily available---these fuzzers randomly sample syntax trees. Inputs generated in this way are guaranteed to be \emph{syntactically valid}. By enforcing bounds on the expansion of recursive production rules and other repeating elements, the size of generated test inputs can also be controlled. In Section~\ref{sec:bgf}, we provide an algorithm for sampling size-bounded test inputs from a context-free grammar provided in an extended BNF notation.

Although grammar fuzzing produces \emph{syntactically valid} test inputs by construction, generating inputs that are \emph{semantically valid} is challenging. For example, we empirically found that the probability of a randomly sampled ChocoPy program of size 3 (precisely defined in Section~\ref{sec:bgf}) being semantically valid is less than 9\%. 

\subsubsection{Semantic Fuzzing}

Recently developed tools such as Zest~\cite{Padhye19-zest},  Nautilus~\cite{Aschermann19}, and Superion~\cite{Wang19} combine structure-aware (e.g. grammar-based) input generators with code coverage feedback. The hope is that such feedback will help generate inputs that are not only syntactically valid, but also exercise various code paths in the compiler corresponding to semantic checks. In fact, Zest is specifically designed to generate \emph{semantically valid} inputs for programs such as compilers. We therefore found Zest a very promising approach for generating a test corpus for ChocoPy.

While Zest-produced test suites were comprehensive---achieving about 95\% line coverage on the ChocoPy type-checker---the generated test corpora were not \emph{concise}. 
For example, the size-bounded Zest-generated program in Fig.~\ref{fig:chocopy-cbgf-ex} simultaneously achieves novel coverage related to the handling of \code{while} loops, \code{for} loops, and \code{if-else} expressions. However, the program also contains certain \emph{redundant} features---those that exist in other inputs in the corpus---such as \code{pass} statements, assignments, and list indexing. This is sometimes referred to as \emph{collateral coverage} in the literature~\cite{Harman10}. We prefer not to provide such a compound input to undergraduate students developing a compiler, as (1) it does not immediately suggest an implementation goal and (2) it is not ideal for debugging failures.

\subsection{Test-Case Reduction}
\label{sec:related-reduction}

A natural solution to the conciseness problem presented by Zest-generated inputs is to simply minimize them. In general, finding a minimal input that exhibits a given behavior (e.g. triggers a bug, or exercises certain program features), is an NP-hard problem. Starting with an initial input of size $n$, there are $\mathcal{O}(2^n)$ possible subsets of the starting input itself, not to mention other small inputs that contain elements not present in the initial input.

Techniques such as Delta Debugging (DD)~\cite{Zeller02} find locally minimal inputs that are subsets of the initial input in worst-case $\mathcal{O}(n^2)$ steps. One drawback of DD applied on the string representation of inputs is that deleting individual characters and contiguous substrings often results in inputs that have invalid syntax; therefore, most subsets do not exhibit the desired behavior.  Hierarchical Delta-Debugging (HDD)~\cite{Misherghi06} solves this problem by applying a DD-like algorithm on a tree representation of parsed inputs. HDD requires knowledge of the input syntax, which is readily available in our application. Similarly, Perses~\cite{Sun18} utilizes a grammar to perform reductions and guarantees that each reduction step also produces a syntactically valid program.

We used state-of-the-art implementations of DD and HDD developed by Hodovan et al.~\cite{Hodovan16-a, Hodovan16-b, Hodovan17-c, Hodovan17-d} on Zest-generated ChocoPy programs. Fig.~\ref{fig:chocopy-cbgf-ex-reduced} depicts a minimized version of the program listed in Fig.~\ref{fig:chocopy-cbgf-ex}, where the reduction criterion was that the reduced input achieves at least the unique same coverage as achieved by the original input. The minimization takes about 30 seconds to run, and achieves a 50\% reduction in test case size---the redundant \code{pass}, assignment, etc. has been removed. However, Fig.~\ref{fig:chocopy-cbgf-ex-reduced} still contains multiple loops, branching statements, etc. In the next section, we will describe a novel solution that produces inputs that are much more concise, \emph{for free}.

\begin{figure}[t]
    \begin{minted}
    [
    framesep=2mm,
    baselinestretch=1.2,
    fontsize=\footnotesize,
    xleftmargin=5mm,
    linenos
    ]   
    {python}
while not (0):
    for a in a:
        b and True

    (0).a = (c)[None if c else 1 if a else ""] 
    pass
    \end{minted}
    \caption{ChocoPy Program generated using coverage-guided bounded grammar-based fuzzing with size bounds of $(3, 3, 3)$.}
    \label{fig:chocopy-cbgf-ex}
\end{figure}

\begin{figure}[t]
    \begin{minted}
    [
    framesep=2mm,
    baselinestretch=1.2,
    fontsize=\footnotesize,
    xleftmargin=5mm,
    linenos
    ]   
    {python}
while A:
    for A in "":
        A= None if A if None else A else A 
    \end{minted}
    \caption{Minimized ChocoPy program achieving the same novel coverage as achieved by the program in Fig. \ref{fig:chocopy-cbgf-ex}.}
    \label{fig:chocopy-cbgf-ex-reduced}
    \vspace{-0.5cm}
\end{figure} \section{\TECH{}}
\label{sec:proposed}

Our proposed technique leverages the scalability advantages of grammar-based coverage-guided fuzzing while avoiding the constraints of the fuzz-then-reduce approach. The \emph{key idea} in our approach is to grow a test corpus \emph{bottom-up} by (1) using coverage-guided bounded grammar fuzzing (CBGF) to generate small inputs \emph{by construction} and (2) iteratively increasing the input size, inspired by iterative-deepening-based search algorithms~\cite{Korf85}. We call our approach \emph{\tech{}}. 

Figs.~\ref{fig:bonsai-ex1}, \ref{fig:bonsai-ex2}, and \ref{fig:bonsai-ex3} show a total of eleven ChocoPy programs saved during various rounds of \tech{} (comments added manually). These programs are concise and the language features they exercise can be easily discerned. In our opinion, they look almost like hand-written test cases that are precisely designed for testing specific features of the ChocoPy language semantics. However, they were generated completely automatically and without knowledge of any typing rules. We next build a series of concepts leading up to a description of the \tech{} algorithm.

\begin{figure}[t]
    \begin{minted}
    [
    framesep=2mm,
    baselinestretch=1.2,
    fontsize=\footnotesize,
    ]   
    {python}
# (Ex. A) Single pass statement
pass

# (Ex. B) Simple assignment statement
a:object = 1

# (Ex. C) Function definition with return
def a():
    return
    \end{minted}
    \caption{Three examples of ChocoPy programs saved  during \tech{}, in a corpus produced by $\mathcal{F}_{1,1,1}$.}
    \label{fig:bonsai-ex1}
\end{figure}

\begin{figure}[t]
    \begin{minted}
    [
    framesep=2mm,
    baselinestretch=1.2,
    fontsize=\footnotesize,
    ]   
    {python}
# (Ex. D) Class definition with attribute declaration
class a(object):
    a:int = 1
pass

# (Ex. E) Indexing into a string
("a")[0]

# (Ex. F) Less-than comparison on two integers
0 < 0

# (Ex. G) Equality comparison on two strings
"" == "a"

# (Ex. H) Function definition with two arguments
def a(b:str, a:int):
    pass
    
    \end{minted}
    \caption{Four examples of ChocoPy programs saved during \tech{} by  $\mathcal{F}_{2,1,1}$,  $\mathcal{F}_{1,2,1}$, and  $\mathcal{F}_{1,1,2}$.}
    \label{fig:bonsai-ex2}
    \vspace{-0.2cm}
\end{figure}

\begin{figure}[t]
    \begin{minted}
    [
    framesep=2mm,
    baselinestretch=1.2,
    fontsize=\footnotesize,
    ]   
    {python}
# (Ex. I) Nested list expression
[[1], [None]]

# (Ex. J) Object construction and attribute assignment
class a(object):
    a:int = 1
(a()).a = 1

# (Ex. K) Nested functions
def a():
    def b():
        pass
    return
    \end{minted}
    \caption{Example ChocoPy programs saved in the \tech{} corpus of  $\mathcal{F}_{3,3,3}$.}
    \label{fig:bonsai-ex3}
\vspace{-0.2cm}
\end{figure}

\subsection{Bounded Grammar Fuzzers}
\label{sec:bgf}

We start by considering an input generator that can randomly sample inputs of a bounded size, where the bounds are based on the definition of an input language's grammar. 
We can observe three properties of a ChocoPy program to get an idea of how we might bound the input space.
\begin{enumerate}
    \item \textit{idents}: the number of new unique identifiers (variable names, function names, class names) excluding predefined identifiers (e.g. \code{int}).
    \item \textit{items}: the maximum number of elements in a linear group. This can correspond to the maximum number of statements in a block, arguments  in a function definition, arguments in a list expression, etc.
    \item \textit{depth}: the maximum number of times an expression, statement, or function definition is nested.
\end{enumerate} 

For the ChocoPy example in Fig. \ref{fig:chocopy-ex1}, we have \emph{idents}=4 (\code{is\_zero, items, idx, val}), \emph{items}=5 (comma-separated list elements on line 6), and \emph{depth}=3 (triply nested expressions on line 6). 
For the example in Fig. \ref{fig:chocopy-ex2}, we have \emph{idents}=7 (\code{A, setx, equals, self, x, y, a}), \emph{items}=4 (top-level statements in the program), and \emph{depth}=2 (doubly nested \code{if} statements on lines 9--11).

We can bound the input space if we restrict the maximum value of \emph{idents}, \emph{items}, and \emph{depth} for any generated ChocoPy program. 
We will now generalize this to any language.

Consider a specification for the syntax of an input language in the form of a context-free grammar $\mathcal{G}$. We consider definitions in an extended Backus–Naur form~\cite{EBNF96}, where $\mathcal{G}$ consists of a set of terminals $\mathcal{T}$, a set of non-terminals $\mathcal{N}$, a start symbol $S \in \mathcal{N}$, and a set of production rules of the form:
$$ A \longrightarrow \alpha, \quad\text{where $A \in \mathcal{N}$ and $\alpha = a_1 a_2 \ldots $}$$
\noindent The right-hand side of production rules $\alpha$ are a sequence of zero or more \emph{symbols} which are defined recursively as follows: a \emph{symbol} is either a terminal in $\mathcal{T}$, a non-terminal in $\mathcal{N}$ or of the form $[b]^*$, where $b$ is a symbol. The Kleene-star in the final form has the usual meaning and enables non-recursive definitions of linear repeating sequences, e.g. list of statements or arguments to a function call. We also consider a special class of terminals $\tau \subseteq \mathcal{T}$ whose concrete values are user-defined (e.g. identifiers) instead of predefined (e.g. `\code{+}' or `\code{while}'). In the ChocoPy grammar---included in our online repository (ref.~ Section~\ref{sec:intro})---we have $\tau = \{ \mathit{ID}, \mathit{IDSTRING} \}$.

Now consider the set of programs $\mathcal{P} = \{p : p \sim \mathcal{G}\}$. Each program $p$ has a corresponding derivation tree $t$ from $\mathcal{G}$. We are interested in bounding the following properties:

\begin{enumerate}
    \item $\textit{idents}(p)$: The maximum number of distinct values for any terminal in $\tau$ (e.g. number of distinct identifiers) observed across the entire tree $t$.
    \item $\textit{items}(p)$: The maximum number of repetitions in any expansion of a Kleene-star (e.g. number of statements in a block) when generating $t$.
    \item $\textit{depth}(p)$: The maximum number of expansions of the same non-terminal (e.g. \code{expr}) in any path from the root to any leaf node in $t$.
\end{enumerate}
We can then define a smaller input space $\mathcal{P}_{m,n,d}$, where 
\begin{equation*}
    \mathcal{P}_{m,n,d} = \left\{p \quad :
    \begin{aligned}\quad & p \in \mathcal{P}\\ &\text{idents}(p) \le m,\\ &\text{items}(p) \le n,\\ &\text{depth}(p) \le d
    \end{aligned}
    \right\}
\end{equation*}

\noindent For example, the ChocoPy program in Fig. \ref{fig:chocopy-ex1} belongs to $\mathrm{ChocoPy}_{4,5,3}$, but the program in Fig.~\ref{fig:chocopy-ex2} does not. Both of them belong to $\mathrm{ChocoPy}_{7, 5, 3}$. Neither is in $\mathrm{ChocoPy}_{1, 1, 1}$.

\begin{algorithm}[t]
\caption{Bounded grammar sampling algorithm. \newline $\mathcal{G}$ is a grammar; $m$, $n$, and $d$ are positive integers.} \label{sampling-algo}
\begin{algorithmic}
\small
\Function{BoundedSample}{$\mathcal{G}$, symbol $a$, $m$, $n$, $d$} 
    \State \textbf{case} typeof($a$):
        \State \quad terminal $t$: \Return \textproc{concretize}($t, m$) \Comment See text...
        \State \quad repetition $[b]^*$: \Return concatenate(
        \State \qquad \qquad \qquad \quad [\textproc{BoundedSample}($b, m, n, d$)
        \State \qquad \qquad \qquad \qquad for $i \in \{0\ldots $ chooseRandom($[0 \dots n]$)$\}$])
        \State \quad nonterminal $A$: \Return 
        \State \qquad \qquad \qquad \qquad \quad \textproc{sampleNonTerminal}($\mathcal{G}, A, m, n, d$)
\EndFunction
\Function{sampleNonTerminal}{$\mathcal{G}$, nonterminal $A$, $m$, $n$, $d$}
    \If{$\mid$\textproc{NT\_Expansions}($\mathcal{G}, A)\mid == 0$}
        \State $p \gets 1$ \Comment{Expand to leaf node}
    \ElsIf{$\mid$\textproc{T\_Expansions}($\mathcal{G}, A)\mid == 0$}
        \State $p \gets 0$ \Comment{Expand to non-leaf node}
    \Else 
        \State Let $c \gets $ number of expansions of $A$ from root to here
\State $p \gets (c + 1) / (d + 1)$ \Comment{Probability of leaf expansion}
    \EndIf
    \State \textbf{with} probability $p$
    \State \qquad Let $A \rightarrow \alpha$ = chooseRandom(\textproc{T\_Expansions}($A$))
    \State otherwise
    \State \qquad Let $A \rightarrow \alpha$ = chooseRandom(\textproc{NT\_Expansions}($A$))
    \State \Return concatenate([\textproc{sample}($b, m, n, d$) for $b$ in $\alpha$]
\EndFunction
\Function{T\_Expansions}{$\mathcal{G}$, nonterminal $A$}
    \State \Return all expansions $A \rightarrow \alpha$ in $\mathcal{G}$ where 
    \State \qquad \qquad \qquad $\forall a_i \in \alpha$, typeof($a_i$) == terminal
\EndFunction
\Function{NT\_Expansions}{$\mathcal{G}$, nonterminal $A$}
        \State \Return all expansions $A \rightarrow \alpha$ in $\mathcal{G}$ where 
    \State \qquad \qquad \qquad $\exists a_i \in \alpha:$ typeof($a_i$) == nonterminal
\EndFunction
\end{algorithmic}
\end{algorithm}

Algorithm \ref{sampling-algo} details the procedure we use for sampling programs in $\mathcal{P}_{m,n,d}$. The parameters to function \textproc{BoundedSample} are a grammar $\mathcal{G}$, a symbol $a$, and bounds $m, n, d$; the function returns a string which is an expansion of symbol $a$ that obeys the provided bounds. A top-level call to \textproc{BoundedSample} with $a = S$, the start symbol of the grammar, produces a random program in $\mathcal{P}_{m,n,d}$. 

The sampling algorithm has a similar structure to the PTC1 grammar-sampling procedure described by Luke~\cite{Luke00}; the following discussion clarifies specific algorithmic details.

In general, since $a$ can be any type of symbol---terminal, nonterminal, or a group with Kleene-star--- \textproc{BoundedSample} performs different logic depending on the type of $a$. 
\begin{enumerate}
    \item When $a$ is a terminal symbol, it is concretized as follows: If $a \in \tau$, then one of $m$ pre-populated expansions is uniformly chosen at random (e.g. if the terminal represents an identifier, then one of say \texttt{a\_1}, \texttt{a\_2}, \dots, \texttt{a\_m} is returned uniformly at random). Otherwise, $a$ has exactly one concrete value (e.g. `\texttt{+}' or `\texttt{while}'), which is returned directly.
    \item If $a$ is a repetition $[b]^*$, we choose a number of expansions $i$ uniformly at random in the range $[0, n]$. Then, we recursively call \textproc{BoundedSample} with symbol $b$ for $i$ times and the results are concatenated.
    \item If $a$ is a nonterminal $A$, then with a calculated probability $p$ we return the output of \textproc{BoundedSample} on a randomly chosen terminal expansion. Otherwise, we use a randomly chosen nonterminal expansion.
The probability $p$ is a function of the number of times $A$ has been expanded from the root and the maximum depth parameter $d$.  It ensures that the program cannot have a depth larger than $d$, while favoring nonterminal expansions when the nesting depth is relatively smaller. The calculation of $p$ differs from that used by Luke in PTC1~\cite{Luke00}, since we are interested in bounding the maximum nesting along any given path in a derivation tree, instead of bounding the size of the tree itself.\end{enumerate}

\subsubsection*{Preliminary Results with ChocoPy} 

 There is a natural dichotomy between \emph{conciseness} and \emph{comprehensiveness}. Tiny bounds such as $(1, 1, 1)$ produce very concise inputs, but they do not exercise many language features. Additionally, most randomly sampled inputs of size $(1, 1, 1)$ are well-typed. As we increase the bounds, the likelihood of a randomly sampled program being semantically valid diminish.

\begin{figure*}[t]
    \includegraphics[scale=0.42]{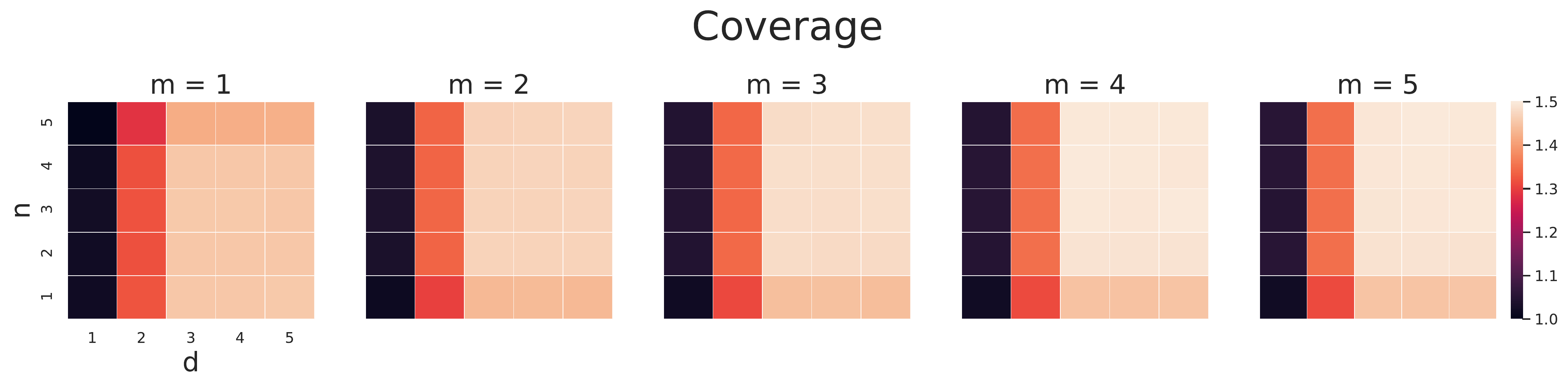}
    \includegraphics[scale=0.42]{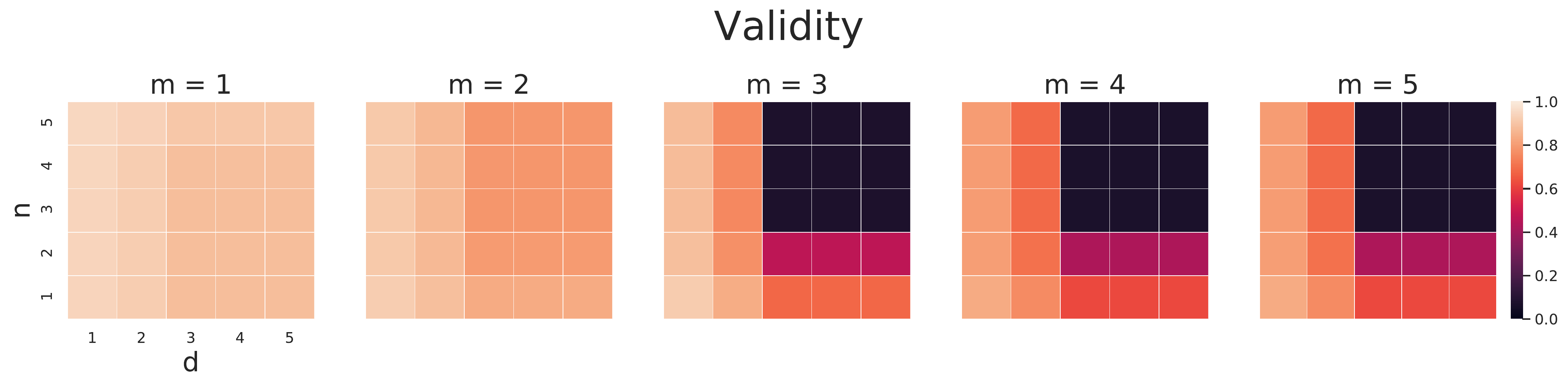}
    \caption{Properties of ChocoPy programs randomly sampled using \textproc{BoundedSample} with various size bounds for 3 hours each. Data shows normalized \emph{branch coverage} and fraction of \emph{semantically valid} programs in sampled population (average over 10 repetitions per config.). Higher values are better.}
    \label{fig:chocopy-heatmaps}
\end{figure*}

For preliminary experiments, we ran small 3-hour fuzzing sessions using bounded grammar sampling for all configurations where $m$, $n$, and $d$ were between 1 and 5 each---a total of 125 configurations. Each experiment was repeated ten times to account for randomness. We then measured (1) \emph{branch coverage} in the ChocoPy reference typechecker across all the inputs generated during each experiment, and (2) fraction of generated inputs that were \emph{semantically valid}. Fig.~\ref{fig:chocopy-heatmaps} shows averages of the fraction validity and relative branch coverage for all 125 configurations. We noticed that bounds such as $(3, 3, 3)$ were able to achieve high coverage; however, the fraction of valid inputs generated for  was concerning (only 9\%). We next consider a feedback-directed variant of the bounded grammar sampling fuzzer that can produce inputs that are more likely to be semantically valid.

\subsection{Coverage-Guided Bounded Grammar Fuzzing (CBGF)}
\label{sec:cbgf}

In order to incorporate a feedback from test execution, we enhance our bounded grammar sampling technique to a \emph{coverage-guided} bounded grammar fuzzer (CBGF). Algorithm~\ref{zest-algo} describes CBGF. It is almost a standard coverage-guided fuzzing loop (e.g. as described by B{\"o}hme et al.~\cite{Bohme17}), but focuses on generating a comprehensive test-case corpus rather than discovering program crashes\footnote{We assume that the reference program being analyzed is not buggy. If we find any crashes, we apply a patch and restart from the beginning.}. The technique expects an instrumented version of the test program, such as the ChocoPy reference compiler; the instrumentation provides a way to receive feedback (e.g. code coverage) from test execution. Test execution on a given input can also return additional feedback such as whether the input was semantically valid or not (e.g. based on whether type-checking succeeded or if there were any errors). The function $\textproc{CBGF}$ is given an ordered set of initial \emph{seed inputs} in $\mathcal{S}$. The main fuzzing loop continuously cycles through the set $\mathcal{S}$, picking each input in order (sometimes with repetition to increase energy~\cite{Bohme17}), mutating it, and executing the test program with the mutated input to receive feedback. If the feedback is \emph{interesting} (e.g. coverage includes a program location that is not exercised by any other input in $\mathcal{S}$ so far), then the mutated input is added to $\mathcal{S}$. The loop ends after a fixed time budget, and the resulting corpus of inputs $\mathcal{S}$ is returned.

The two main unspecified components in this algorithm are how $\textproc{Mutate}$ works (Line~\ref{line:zest-algo-mutate}) and what the interestingness criteria is for saving new inputs  (Line~\ref{line:zest-algo-interesting}). We use an off-the-shelf implementation of Zest~\cite{Padhye19-zest}, a structure-aware coverage-guided fuzzer that is well suited for our application\footnote{We elide details of all other heuristics in Algorithm~\ref{zest-algo}, since we inherit them from the original Zest implementation~\cite{Padhye19-zest}. The search heuristics are not important to our proposed technique, which works at a higher level.}. In Zest, all inputs---including the initial seed inputs---are generated using some sampling procedure called a \emph{generator}; in our case, the generator is simply the bounded grammar sampler (ref. Algorithm~\ref{sampling-algo}). Each input is associated with a sequence of pseudo-random \emph{choices} made during the sampling procedure, which uniquely determine the input produced by that procedure. In Algorithm~\ref{sampling-algo}, this includes the ``random'' choices made in expanding production rules and concretizing terminal values. Zest records these choices in a vector, which is associated with the corresponding $input$. The \textproc{Mutate} function in Algorithm~\ref{zest-algo} works by performing random point mutations on these recorded pseudo-random choices and then replaying $\textproc{BoundedSample}$ with the specified choices and with the given bounds to produce $input'$. Essentially,  $\textproc{BoundedSample}$ is implicitly parameterized by the source of pseudo-randomness, which Zest controls---in the implementation, by simply overriding \code{java.util.Random}. We expect the returned value $input'$ to be a syntactically valid input that is subtly different from---that is, a structural mutation of---the original input $input$~\cite{Padhye19-zest}. Note that if $input$ was a member of the initial set of seeds, then the size bounds $(m,n,d)$ provided to $\textproc{Mutate}$ may be \emph{larger} than the bounds used to originally generate $input$; we will exploit this fact in the Section~\ref{sec:bonsai}. 

The criteria used by Zest to determine whether to save $input'$ (Line~\ref{line:zest-algo-interesting} in Algorithm~\ref{zest-algo}) is the following: the feedback from execution of $p$ on $input'$ is interesting if (1) there is new code coverage, regardless of the validity of $input'$, or (2) $input'$ is semantically valid and it achieves new coverage when compared to all other semantically valid inputs in $\mathcal{S}$. Zest thus favors saving semantically valid inputs. Section~\ref{sec:restricted} describes a tweak to this criterion we make in some scenarios.

\subsubsection*{The $\mathcal{F}$ notation}

We now define some short-hand notation that will be useful when describing our proposed \tech{} technique. Let $\mathcal{F}^{\mathcal{G}, p}_{m, n, d}$ denote a coverage-guided bounded grammar fuzzer (CBGF) parameterized by grammar $\mathcal{G}$, test program $p$, size bounds $m$, $n$, and $d$. As per Algorithm~\ref{zest-algo}, $\mathcal{F}^{\mathcal{G}, p}_{m, n, d}$ is a function that accepts an ordered set of inputs and returns a corpus of the same type. Since the grammar and target program are usually fixed in a given application, we will omit the superscripts hereon; therefore, $\mathcal{F}_{m,n,d}$ is a CBGF of size bounds $(m,n,d)$.

\subsubsection*{Preliminary Results with ChocoPy} 

As described in Section~\ref{sec:related-reduction}, simply using Zest followed by input minimization on the resulting corpus still lacks conciseness. The program in Fig.~\ref{fig:chocopy-cbgf-ex} was produced using $\mathcal{F}_{3,3,3}$ in seedless mode~\cite{You19}. The program in  Fig.~\ref{fig:chocopy-cbgf-ex-reduced} is its corresponding reduction after applying hierarchical delta debugging~\cite{Misherghi06}---the invariant being that the reduced input still meets the same interestingness criteria from Algorithm~\ref{zest-algo}. A full HDD-reduced corpus of Zest-generated ChocoPy programs can be found in our online repository (ref.~Section~\ref{sec:intro}).

\begin{algorithm}[t]
\caption{Coverage-Guided Bounded Grammar Fuzzing}
\label{zest-algo}
\begin{algorithmic}[1]
\small
\Require Instrumented program $p$, Grammar $\mathcal{G}$, Bounds $m$, $n$, $d$
\Function{CBGF}{Seed inputs $\mathcal{S}$} \Comment Returns corpus $\supseteq       \mathcal{S}$    

\Repeat 
        \State $\textit{input} \gets$ next($\mathcal{S}$) \Comment{Cycle through $\mathcal{S}$}
        \State $\textit{input'} \gets \textproc{Mutate}(\textit{input}, \mathcal{G}, m, n, d)$  \label{line:zest-algo-mutate} \Comment See text...
        \State $\textit{feedback} \gets \textproc{Execute}(p, input')$ \Comment{validity + coverage}
        \If{\textit{feedback} is interesting} \Comment{ new coverage?} \label{line:zest-algo-interesting} 
            \State $\mathcal{S} \gets \mathcal{S} \cup \textit{input'}$
        \EndIf
\Until{time budget expires}
    \State \Return $\mathcal{S}$ 
\EndFunction
\end{algorithmic}
\end{algorithm}

\subsection{\TECH{}}
\label{sec:bonsai}

Our novel solution is to build a concise test corpus from the bottom up by using a set of CBGFs with gradually increasing size bounds. The intuition is that the smaller CBGFs would initially build a corpus of tiny test corpus covering simple features, and larger CBGFs can build on the smaller programs to generate more complex test cases that achieve better coverage. By inceasing the size bounds gradually at each step, we expect the complex test cases in later stages to be structural mutations of test cases discovered in earlier stages; thus, we hope to simultaneously achieve validity, conciseness, and comprehensiveness. We now define a way to iteratively increment the size of a CBGF, which allows us to create a formal procedure for this approach.  

Given upper bounds $M$, $N$, and $D$, we can consider the set of CBGFs
\begin{equation*}
    \mathcal{C}_{M,N,D} = \left\{\mathcal{F}_{m, n, d} \quad :
    \begin{aligned}\quad & 1 \le m \le M \\ &1 \le n \le N \\ &1 \le d \le D
    \end{aligned}
    \right\}
\end{equation*}
With upper bounds $(3, 3, 3)$, we would have $27$ different CBGFs in set $\mathcal{C}_{3,3,3}$. 

We define a \emph{partial order} $\le$ over $\mathcal{C}_{M,N,D}$ as follows: \begin{align*}
    \mathcal{F}_{m, n, d} \le \mathcal{F}_{m', n', d'} \quad \Longleftrightarrow \quad  m \le m', n \le n',d \le d' 
\end{align*}
\noindent Consequently, $\mathcal{F}_{m, n, d} < \mathcal{F}_{m', n', d'}$ iff $\mathcal{F}_{m, n, d} \le \mathcal{F}_{m', n', d'}$ and $\mathcal{F}_{m, n, d} \neq \mathcal{F}_{m', n', d'}$.

This ordering suggests that $\mathcal{C}_{M,N,D}$ is a lattice with $\mathcal{F}_{1,1,1}$ being the bottom element (denoted $\mathcal{F}_\bot$) and $\mathcal{F}_{M,N,D}$ being the top element (denoted $\mathcal{F}^\top$). Fig.~\ref{fig:lattice} visualizes the lattice for $\mathcal{C}_{2,2,2}$, where the partial order corresponds to graph reachability. In this example, $\mathcal{F}^\top = \mathcal{F}_{2,2,2}$.

Additionally, we define the terms \emph{successor} and \emph{predecessor} with their usual meaning:
\begin{enumerate}
    \item $\mathcal{F}_s$ is a \emph{successor} of $\mathcal{F}$ if $\mathcal{F} < \mathcal{F}_s$ and there exists no CBGF $\mathcal{F}'_s$ such that $\mathcal{F} < \mathcal{F}'_s < \mathcal{F}_s$.
    \item $\mathcal{F}_p$ is a \emph{predecessor} of $\mathcal{F}$ if $\mathcal{F}_p < \mathcal{F}$ and there exists no CBGF $\mathcal{F}'_p$ such that $\mathcal{F}_p < \mathcal{F}'_p < \mathcal{F}$. 
\end{enumerate}
For example, $\mathcal{F}_{2,1,1}$ is a successor of $\mathcal{F}_{1,1,1}$, whereas $\mathcal{F}_{2,2,1}$ is not. Conversely, $\mathcal{F}_{1,1,1}$ is a predecessor of $\mathcal{F}_{2,1,1}$ but not a predecessor of $\mathcal{F}_{2,2,1}$. In Fig.~\ref{fig:lattice}, every node has incoming edges from its predecessors and outgoing edges to its successors. Naturally,  $\mathit{predecessors}(\mathcal{F}_{\bot}) = \mathit{successors}(\mathcal{F}^{\top}) = \{\}$.

We now formally define \emph{\tech{}} as a procedure that begins with the smallest configuration $\mathcal{F}_\bot$ and iteratively increases the size until a given upper bound is reached. 

\begin{algorithm}[t]
\small
\caption{\Tech{} algorithm}\label{bonsai-algo}
\begin{algorithmic}[1]
\Procedure{BonsaiFuzzing}{} 
    \State $\mathcal{F} \gets \mathcal{F}_{\bot}$
        \State
        $\mathit{seeds} \gets \left[\text{random}()\right]$ \Comment{Single random seed} 
        \State
        $\mathit{corpus}(\mathcal{F}_{\bot}$) = $\mathcal{F}(\mathit{seeds})$  \Comment{Run CBGF to generate corpus}
    \State $\mathit{worklist} \gets \mathit{successors}(\mathcal{F})$
    \While{$\mathit{worklist}$ is not empty}
        \For{each $\mathcal{F} \text{ in } \mathit{worklist}$} \Comment{Parallelizable}
        
            \State $P \gets \mathit{predecessors}(\mathcal{F})$
            \State $\mathit{seeds} \gets \textproc{SortBySize}\left(\bigcup_{\mathcal{F}_p \in P} \mathit{corpus}(\mathcal{F}_p)\right)$
            \State $\mathit{corpus}(\mathcal{F}) \gets \mathcal{F}(\mathit{seeds})$  \Comment{Run CBGF}
        \EndFor
        \State $\mathit{worklist} \gets \bigcup_{\mathcal{F}_s \in \mathit{worklist}} \mathit{successors}(\mathcal{F}_s)$
    \EndWhile
    \State \Return $\mathit{corpus}(\mathcal{F}^{\top}$)
\EndProcedure
\end{algorithmic}
\end{algorithm}

Algorithm~\ref{bonsai-algo} describes the procedure for \tech{}. Variable $\mathcal{F}$ is initialized to the smallest CBGF $\mathcal{F}_{\bot}$. Recall from Algorithm~\ref{zest-algo} that a CBGF is a function that is given a set of seed inputs and returns a test corpus. Initially, we have no seeds. We thus start by running the CBGF $\mathcal{F}_{\bot}$ with one random seed input (similar to SLF~\cite{You19}) to produce $\mathit{corpus}(\mathcal{F}_{\bot})$. Then, a $\mathit{worklist}$ is populated with the $\mathit{successors}(\mathcal{F})$. For each unprocessed element in the worklist---that is, each unexecuted fuzzer---we prepare its $\mathit{seeds}$ by taking a union of all test cases in the $\mathit{corpus}$ generated by each of its predecessors. The seeds are also sorted by size in ascending order (so that Algorithm~\ref{zest-algo} encounters smaller inputs to mutate first). We then run $\mathcal{F}$, save its resulting $\mathit{corpus}$, and repeat this process. Eventually, we reach the point where $\mathcal{F} = \mathcal{F}^\top$ and there are no more successors. The final corpus is the result of $\mathcal{F}^\top$.  

Consider a sample run of \tech{} over the set $\mathcal{C}_{3,3,3}$. We start by running the CBGF $\mathcal{F}_{\bot} = \mathcal{F}_{1,1,1}$ with one randomly generated seed input. Fig.~\ref{fig:bonsai-ex1} shows three sample test cases saved in the resulting $\mathit{corpus}(\mathcal{F}_{1,1,1})$. We can see that the generated programs are small in size and test simple language features. These inputs will then be used as seeds in successor CBGFs: $\mathcal{F}_{2,1,1}, \mathcal{F}_{1,2,1},$ and $\mathcal{F}_{1,1,2}$. Fig. \ref{fig:bonsai-ex2} lists some programs saved in the corresponding corpora of these fuzzers. We can now start to see programs with slightly complex features, such as class attributes, binary expressions, and functions with multiple parameters. We repeat the process until we reach $\mathcal{F}_{3,3,3}$, the top element of the lattice $\mathcal{C}_{3,3,3}$. 
Fig.~\ref{fig:bonsai-ex3} shows some example programs saved in the its corpus. More complex features such as nested list expressions and nested function definitions are demonstrated in these generated programs. Note that some of these inputs may have been copied verbatim from its seeds, having been discovered by predecessors. The final corpus necessarily incorporates the corpora generated by all CBGFs in the lattice. The full corpus of ChocoPy programs generated using \tech{} can be found in our online repository (ref.~Section~\ref{sec:intro}).

\subsection{\TECH{} with Extended Lattice}
\label{sec:restricted}

So far, we have restricted test generation to only those programs that are semantically valid. By and large, we want semantic rules (e.g. well-typed addition) to be exercised in valid representative programs, as opposed to larger invalid programs that contain these as subexpressions. However, in order to have a comprehensive test corpus, we also need some invalid input programs for testing various error paths in the semantic analysis (e.g. duplicate variable definition, non-boolean condition to \code{while}, and so on). Ideally, we want these invalid programs to be \textit{concise} as well; that is, they are indicative of the particular error path that is being tested. To achieve this goal, we define two variants of CBGF by tweaking the interestingness criterion on  Line~\ref{line:zest-algo-interesting} of Algorithm~\ref{zest-algo}. First, a \emph{restricted}-CBGF is a CBGF that only saves valid inputs: that is, the feedback is considered interesting on Line~\ref{line:zest-algo-interesting} if the input was valid \emph{and} it achieved new code coverage. Second, an \emph{unrestricted}-CBGF saves both valid and invalid inputs, using Zest's interestingness criterion as described in Section~\ref{sec:cbgf}.

We thus add a parameter $v \in \{r, u\}$ to CBGFs, where $r$ denotes restricted and $u$ denotes unrestricted. We use the symbol $\mathscr{F}_{m,n,d,v}$ to denote a CBGF that is parameterized by size bounds as well as the validity restriction (or lack thereof). We can now define a new partial ordering as follows: given two CBGFs $\mathscr{F}_{m, n, d, v}$ and $\mathscr{F}_{m', n', d', v'}$:
\begin{align*}
    \mathscr{F}_{m, n, d, v} \le \mathscr{F}_{m', n', d', v'} \quad &\Longleftrightarrow
    \begin{aligned}\quad & \mathcal{F}_{m, n, d} \le \mathcal{F}_{m', n', d'} \text{ and } \\ 
        & (v = v' \text{ or } v' = u)
    \end{aligned}\end{align*}

The definitions of \emph{successors} and \emph{predecessors} remain the same as before. So we now have
$\mathit{successors}(\mathscr{F}_{1,1,1,r}) = \{\mathscr{F}_{2,1,1,r}, \mathscr{F}_{1,2,1,r}, \mathscr{F}_{1,1,2,r}, \mathscr{F}_{1,1,1,u}\}$. Similarly, we now have $\mathit{predecessors}(\mathscr{F}_{1,2,1,u}) = \{\mathscr{F}_{1,1,1,u}, \mathscr{F}_{1,2,1,r}\}$
The key idea of this lattice is that \textit{restricted}-CBGFs are predecessors of \textit{unrestricted}-CBGFs with the same size bounds. In other words, \textit{unrestricted}-CBGFs with size bounds $m,n,d$ will be able to use as seeds all the valid inputs produced by a fuzzer of the same size bounds, as well as both valid and invalid inputs produced by unrestricted fuzzers with smaller size bounds. The hope is that invalid inputs that are generated by mutating valid inputs are more likely to be concise, as they would trigger fewer semantic errors in a single ChocoPy program.

Setting $\mathscr{F}_{\bot} = \mathscr{F}_{1,1,1,r}$ and $\mathscr{F}^{\top} = \mathscr{F}_{M,N,D,u}$, we can run \tech{} using Algorithm~\ref{bonsai-algo} as is.
 \section{Evaluation}
\label{sec:eval}

\begin{figure*}[ht]
    \centering
    \includegraphics[scale=0.5]{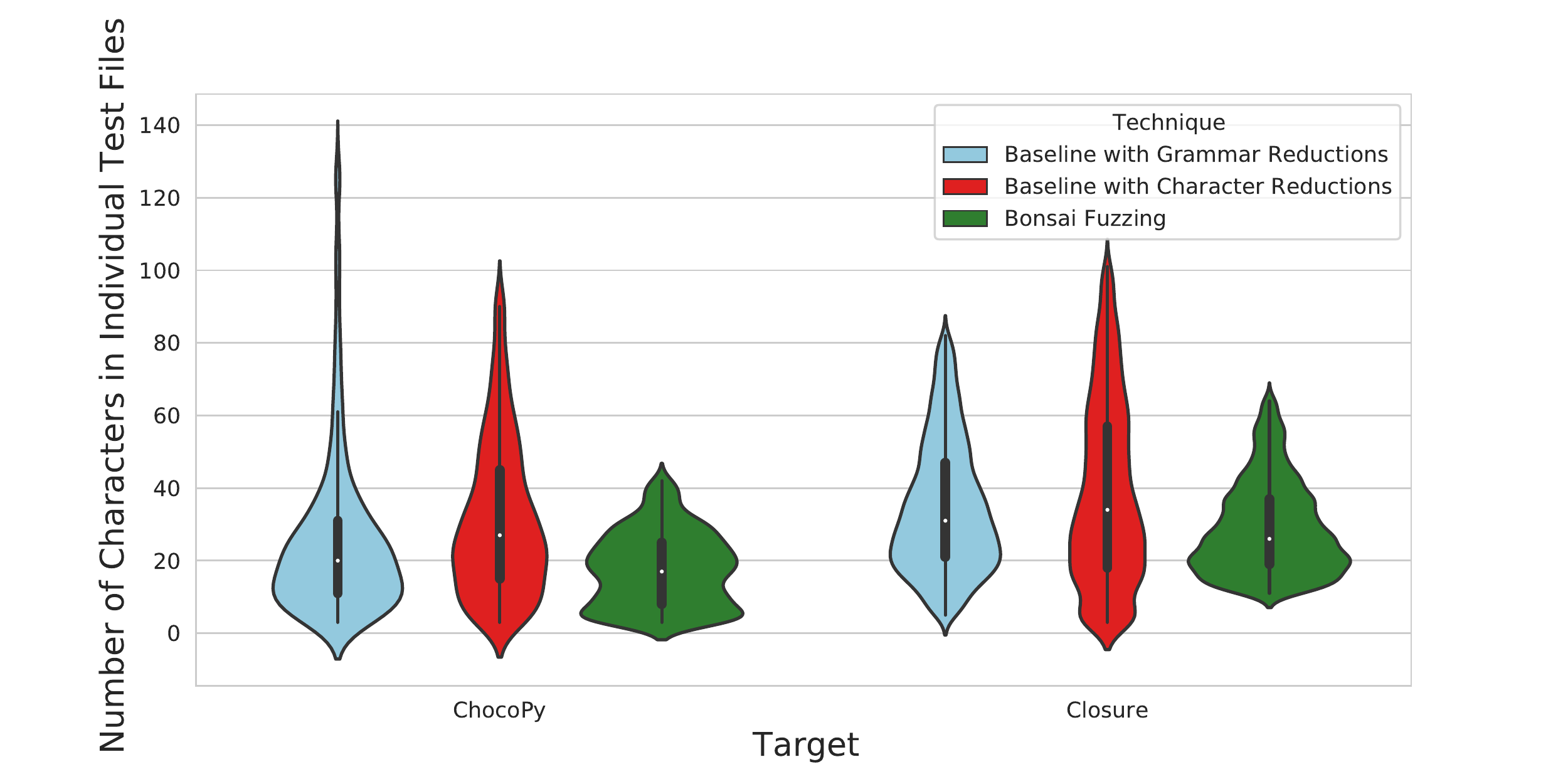}
    \caption{Distribution of size of individual test files, excluding whitespace characters, in saved test corpora. Lower is better.}
    \label{fig:test-size-violin}
\end{figure*}
We evaluate \tech{} by measuring its ability to generate a test corpus containing test cases that are \textit{concise}, \textit{comprehensive}, \textit{semantically valid}, and (where applicable) able to detect faults. We compare \tech{} to a baseline of CBGF (that is; Zest~\cite{Padhye19-zest} with a grammar-based input generator) post-processed with minimization techniques. The baseline is thus the conventional ``fuzz-then-reduce'' approach. We run our evaluation on two test targets: our primary application and a secondary target to ensure that our solution is not biased towards a particular implementation or input language.

\noindent
\begin{enumerate}
    \item ChocoPy~\cite{ChocoPy} reference compiler ($\sim$6K LoC): The test driver reads in a ChocoPy program and runs the semantic analysis / type-checking stage of the ChocoPy reference compiler. For the fault-detection evaluation, we additionally run a differential test on the typed ASTs returned by a reference and buggy compiler (see Section~\ref{sec:eval-mutation}).
    \item Google Closure Compiler\cite{Closure} ($\sim$250K LoC): The test driver (borrowed from prior work~\cite{Padhye19-zest}) expects a JavaScript program as input and performs source-to-source optimizations.
\end{enumerate}

\subsubsection*{Experimental Setup}
\begin{enumerate}
    \item \textbf{Bound}: Overall, we found the bounds of $(M = N = D = 3)$ to be a good trade-off between conciseness and comprehensiveness. We use these bounds for \tech{} as well as for the baseline CBGF.
    \item \textbf{Duration}: We run each CBGF node in the \tech{} extended lattice for one hour, which totals 54 hours of CPU time. We allocate the same 54 hours of CPU time for the baseline CBGF to run\footnote{We chose these durations because one hour is sufficient time for coverage to stagnate for each CBGF node, and because it helps us make a fair comparison with the baseline by fixing total fuzzing duration to a constant. \Tech{} can be optimized by stopping each CBGF early by detecting saturation dynamically, but this would make the total fuzzing duration variable. Our evaluation is conservative.}. 
    \item \textbf{Repetition}: We run each experiment 10 times and report metrics across all repetitions due to the nature of randomness in fuzzing and its effect on results.
\end{enumerate}

\subsubsection*{Minimization Techniques} For the fuzz-then-reduce baseline, we use Picire~\cite{Picire} and Picireny~\cite{Picireny}, which are state-of-the-art~\cite{Hodovan16-a, Hodovan16-b, Hodovan17-c, Hodovan17-d} implementations of character-level~\cite{Zeller02} and grammar-based hierarchical~\cite{Misherghi06} delta debugging respectively. An ``interestingness'' predicate script was required for each of these tools. We provided a predicate that checked whether a candidate minimized input program met the same criterion as was used to save the original input during CBGF (ref.~Line~\ref{line:zest-algo-interesting} in Algorithm~\ref{zest-algo}).
Table~\ref{tab:reduction-times} lists the average CPU-time for each of these reduction tools to minimize an entire corpus.

\subsection{Conciseness: Test Corpus Size}

We evaluate \emph{conciseness} by measuring the size of each test file---excluding whitespace characters---in the generated corpus. Fig. \ref{fig:test-size-violin} displays the distribution of test input sizes for the baseline and \tech{}.

On both targets, we observed that \tech{} produces test files that are statistically significantly lower in size than those of the baseline. The ChocoPy files are on average 42.22\% smaller than the results of grammar-based reduction and 44.51\% smaller than the results of character based reduction. The Closure files are on average 16.49\% smaller than the results of grammar-based reduction and 25.56\% smaller than the results of character-based reduction. We also see that the variance of the size of files in the violin plot of \tech{} is much lower than that of the baseline. One clear advantage is that \tech{} is able to produce these smaller inputs \textit{without requiring any additional post-processing time}. In contrast, the fuzz-then-reduce approach of the baseline can take up to 6 hours for minimization to run. 

As a sanity check, we also report the number of files in the test corpora as shown in Table {\ref{tab:corpus-size}}. The resulting corpus from \tech{} contains about 18\% fewer files in both targets. This shows that \tech{} does not compensate for its smaller test inputs by having a large number of tests.

\begin{table}[t]
    \centering
    \caption{Time to minimize Zest-saved test inputs (minutes, avg $\pm$ stdev).}
    \label{tab:reduction-times}
        \begin{tabular}{ |c||c|c|  }
         \hline
            & ChocoPy & Closure\\
            \hline
            Picireny Grammar Reductions & $56.510 \pm 1.887$ & $356.863 \pm 37.560$ \\
            \hline
            Picire Character Reductions & $20.491 \pm 3.209$ & $392.777 \pm 48.456$ \\
            \hline
        \end{tabular}
\end{table}

\begin{table}[t]
\centering
 \caption{Number of files in test corpus (avg $\pm$ stdev). Lower is better.}
 \label{tab:corpus-size}
        \begin{tabular}{ |c||c|c|  }
         \hline
            & ChocoPy & Closure\\
            \hline
            Baseline & $185.9 \pm 7.666$ & $1507.7 \pm 28.351$ \\
            \hline
            \Tech{} & $152.9 \pm 1.91$2 & $1231.2 \pm 36.705$\\
            \hline
        \end{tabular}
\end{table}

\subsection{Semantic Validity}

One of our goals was to generate a high fraction of semantically valid inputs (ref.~Section~\ref{sec:motivation-problem}). For each input program in the saved test corpora, we re-run the ChocoPy compiler to test whether the input is semantically valid or whether the compiler reports any errors.

\begin{figure}[t]
    \centering
    \includegraphics[scale=0.5]{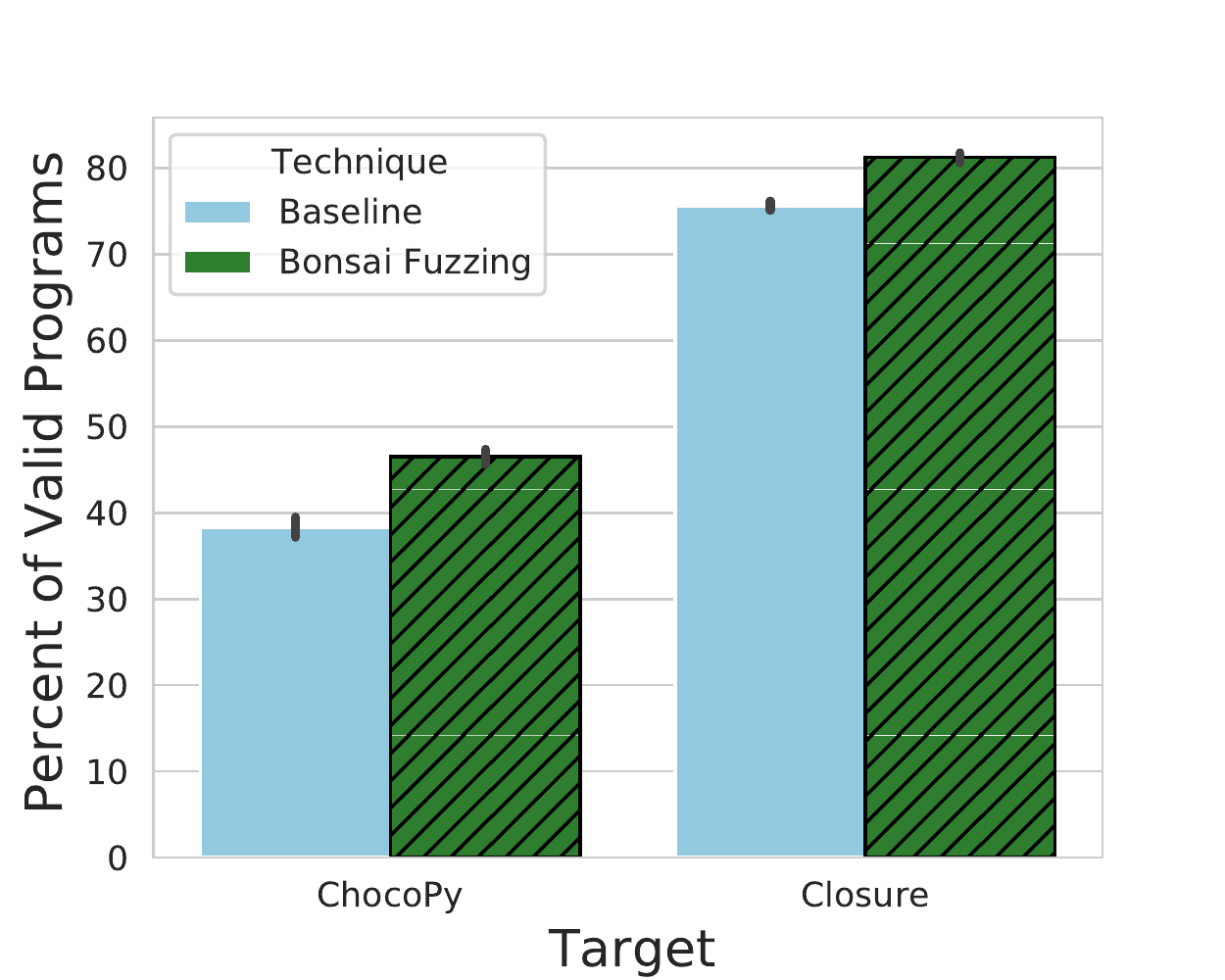}
    \caption{Fraction of semantically valid programs in test corpora (averages with standard deviation). Higher is better.}
    \label{fig:validity}
    \vspace{-0.3cm}
\end{figure}

The average percent of semantically valid programs in the generated corpora is shown in Fig. \ref{fig:validity}. \Tech{} has a statistically significant increase in both targets. On average, it is able to achieve a 21\% improvement in validity in ChocoPy and a 7\% improvement in Closure.  Why is this so? In the initial round of \tech{}, sampling smaller programs leads to a higher likelihood of semantically valid inputs as compared to sampling a larger program from scratch (ref.~Fig.~\ref{fig:chocopy-heatmaps}). In subsequent rounds, it is easier to mutate a small valid program into a slightly larger valid program, as there are less opportunities to introduce errors. We observed that the baseline's seed pool quickly fills up with invalid or large programs early-on in the fuzzing campaign, making it harder to recover in producing diverse valid inputs via random mutations.

We value this improvement in validity resulting from \tech{}, since it means that more language features are being covered by test cases that are semantically valid, which in our opinion results in more meaningful and readable test cases.

\subsection{Comprehensiveness: Coverage}

A key concern when generating small inputs by construction is whether they \emph{comprehensively} exercise various program behaviors as conventional coverage-guided fuzzing.

We measure coverage using a third-party tool: the widely used JaCoCO library\cite{Jacoco}. We report the branch coverage on the semantic analysis classes within each of the benchmarks, similar to approach in \cite{Padhye19-zest}. Since many of the branches are unreachable from our test drivers, it is important to focus on the relative difference between the baseline and \tech{} rather than the raw coverage values.

\begin{figure}[t]
    \centering
    \includegraphics[scale=0.5]{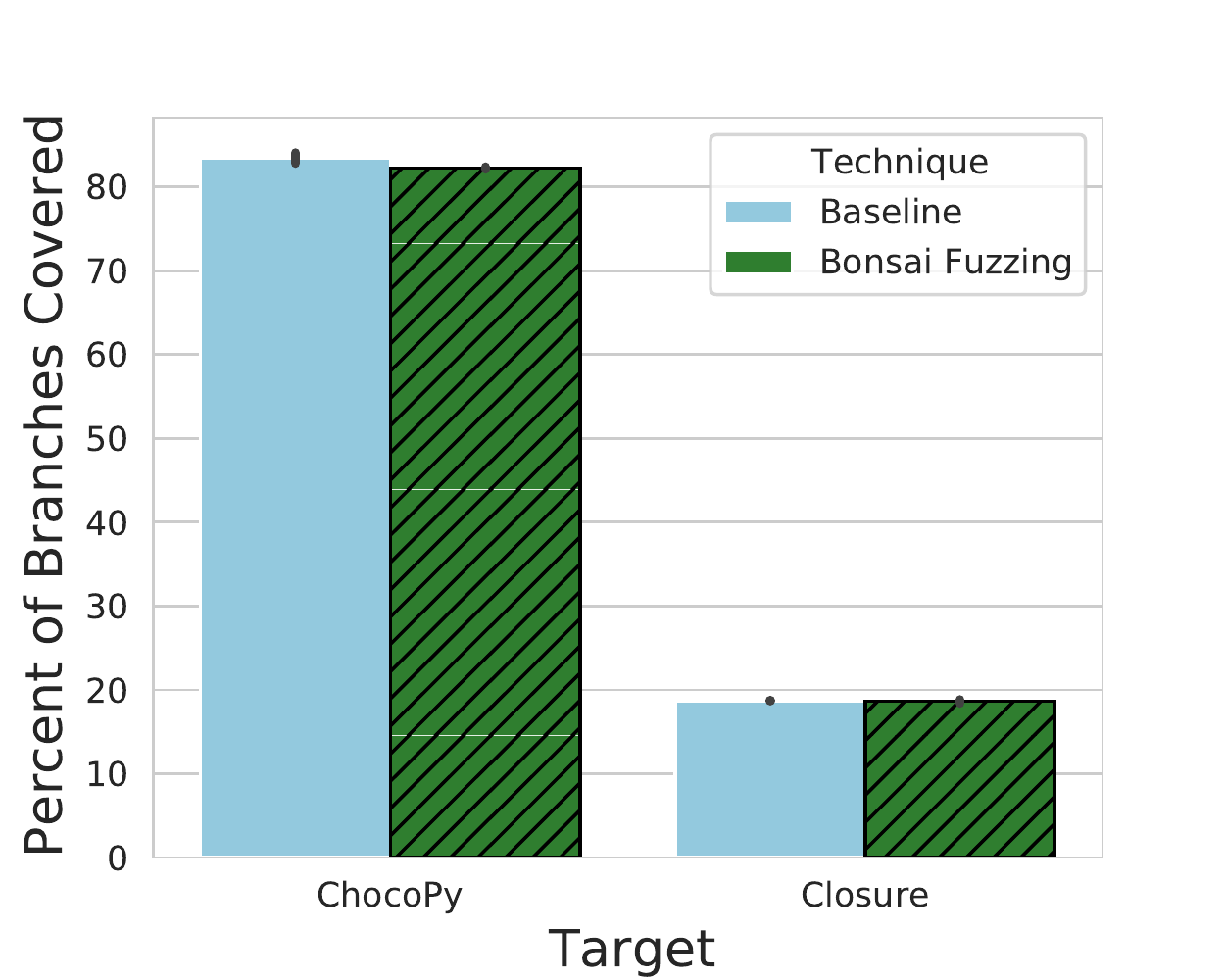}
    \caption{Branch coverage in semantic analysis stages achieved by saved test corpora (averages with standard deviation). Higher is better.}
    \label{fig:branch-cov}
    \vspace{-0.3cm}
\end{figure}

Fig. \ref{fig:branch-cov} shows the branch coverage achieved by the baseline and \tech{} on each of the targets. We can see that both techniques achieve approximately the same branch coverage. On Closure, the difference is statistically insignificant. On ChocoPy, the difference is significant but its effect is small: \tech{} loses $1.175\%$ of branch coverage on average. We are not dismayed with this small reduction. In our application, we can easily incorporate the few test cases from conventional fuzzing that cover logic that is not exercised by \tech{}---in ChocoPy, this is usually just one test case.

\subsection{Fault Detection: Mutation Scores}
\label{sec:eval-mutation}

\begin{table}
    \centering
    \caption{Mutation scores for ChocoPy typechecker (avg $\pm$ stdev)}
    \label{tab:chocopy-mutation-scores}
    \begin{tabular}{ |c||c|  }
     \hline
        Baseline & $91.486 \pm 1.012$\%\\
        \hline
        \TECH{} & $90.428 \pm 0.714$\% \\
        \hline
    \end{tabular}
\vspace{-0.5cm}
\end{table}

Finally, we want to ensure that the concise inputs generated by \tech{} for the ChocoPy target are still useful for catching faults; that is, they can be used for automated grading or providing student feedback. This is essentially a validation of the small scope hypothesis~\cite{Jackson12}.
In a classroom setting, we would compare a candidate buggy student implementation with the reference implementation. For our experimental evaluation, we simulate such a buggy candidate by using a mutation testing tool~\cite{PITest} on a copy of the reference compiler. We run the ChocoPy autograder on the reference compiler and its mutation; if the auto-grader detects a failure, then the mutation is \emph{killed}.

The test corpus saved by \tech{} by itself achieves a mutation-killing score of 81\% on average. This is despite the fact that the fuzzing technique and input-saving criteria is related to coverage improvements within the reference compiler only, and is unaware of program mutations or bugs in student implementations.
As recently observed by Chen et al.~\cite{Chen20}, a better technique for increasing fault detection while minimizing test sizes is to first optimize for coverage and then optimize for mutation scores when coverage saturates. We thus use the corpus produced by \tech{} (and the baseline, for comparison) as seed inputs for a simple grammar-based blackbox fuzzer with the maximum bounds $(3,3,3)$ for 30 minutes. We do this for \emph{each} of the 444 mutated compilers--that is, simulated buggy candidates. If any blackbox -fuzzer-generated input kills the mutation, we say that the corresponding technique kills that mutation.

Table \ref{tab:chocopy-mutation-scores} summarizes these results. Both the baseline and \tech{} achieve more than 90\% mutation-killing score, which we find to be acceptable. We therefore conclude that size-bounded fuzzing does not significantly sacrifice fault detection capability on ChocoPy.
Unfortunately, we cannot report meaningful mutation scores on Closure, since the project does not have a proper differential testing oracle.

 \section{Discussion and Threats to Validity}

Although our original motivation for this work was to synthesize concise test inputs for ChocoPy programming assignments, we also evaluated our technique on the Google Closure Compiler. 
The results of are promising. \Tech{} can synthesize test inputs that are concise by construction, without sacrificing the quality of test inputs in terms of code coverage or mutation scores as compared to the fuzz-then-reduce approach. Moreover, the test inputs produced using \tech{} are smaller in size by 16--45\%. 

However, since the number of target programs we evaluated on is small, we cannot claim that this technique will generalize more broadly. Further, we restricted our evaluation only to compilers, where the input can be represented by a context-free grammar (CFG). We leave the generalization of this technique to other input formats and problem domains as future work.

Further, in our evaluation, we fixed the final size bounds to $(3,3,3)$ and fuzzing duration to one hour per CBGF node. \Tech{} can be improved further by dynamically choosing ideal size bounds and fuzzing duration by monitoring the quality of test inputs saved by each CBGF node.

We were unfortunately unable to test fault detection capabilities of \tech{} on actual student implementations, due to procedural issues with using student-authored assignments for this research. We used \emph{mutation scores} to estimate the ability of \tech{} to catch student bugs. Prior empirical studies have shown this metric to be reasonable~\cite{Just14}, but we cannot make general claims about the impacts of this research in the classroom.

In this paper, we used the notion of \emph{conciseness} of test inputs as a proxy for {readability}, based on what we feel are important features of readable test cases (size and semantic validity). Since our evaluation did not comprise of a user study, we cannot make any subjective claims about human-perceived readability. 
Independently from our work, Roy et al.~\cite{Roy20} have recently worked on improving the readability of automatically generated test \emph{code}, addressing issues such as variable names and code comments.

It has not escaped our notice that the \tech{} technique may also be useful in synthesizing regression tests for fast evolving software. For validating code changes, it is much more efficient to simply run a fixed suite of regression tests than to run a full fuzzing session after every code commit. Concise test inputs, such as those produced using \tech{}, are more likely to be maintainable.

\section*{Acknowledgments}

This research is supported in part by NSF grants CCF-1900968, CCF-1908870, and CNS-1817122, as well as by gifts from Fujitsu and CyLab. The experimental evaluation for this paper was supported by the Amazon AWS Cloud Credits for Research program. We would like to thank Eric Eide for sharing a corpus of C-Reduced~\cite{Regehr12} programs, which helped us form some of our initial insights about program sizes. We also thank the anonymous reviewers for their thoughtful feedback.

\bibliographystyle{IEEEtran}
\bibliography{IEEEabrv,references}
\balance

\end{document}